\documentclass[article,10pt,a5paper,oneside,cyremdash]{ncc}
\FromMargins[h]{2.4cm}{1.6cm}{.8cm}{1.6cm}
\usepackage{cmap}
\usepackage[boldsans]{concmath}
\usepackage{amssymb,fancyvrb,cite,icomma,indentfirst}
\fvset{fontsize=\footnotesize,frame=single,framesep=.4em,
       xleftmargin=0.2em,xrightmargin=0.2em}
\usepackage[utf8]{inputenc}
\usepackage[T1,T2A]{fontenc}
\usepackage[english,russian]{babel}

\usepackage[pdftex,unicode]{hyperref}
\hypersetup{unicode=true,
    pdfauthor={P.V. Moskalev, K.V. Grebennikov, V.V. Shitov},
    pdfsubject={Interval estimation of the mass fractal dimension for anisotropic sampling percolation clusters},
    pdftitle={P.V. Moskalev et al. Interval estimation of the mass fractal dimension for anisotropic sampling percolation clusters},
    pdfkeywords={mathematical modeling, Monte Carlo methods, site percolation, percolation cluster, mass fractal dimension} }

\begin{document}

\begin{titlepage}

\English
\begin{flushleft}\bfseries
UDC 519.676
\end{flushleft}
\begin{flushleft}\bfseries\Large
Interval estimation of the mass fractal dimension for anisotropic sampling percolation clusters
\end{flushleft}
\begin{flushleft}\bfseries
P.V. Moskalev (Voronezh, moskalefff@gmail.com), \\
K.V. Grebennikov, V.V. Shitov (Voronezh, svw@list.ru)
\end{flushleft}
\begin{flushleft}
International conference 
<<Actual problems of applied mathematics, computer science and mechanics>>, \\
Voronezh, 19\--21 September 2011
\end{flushleft}
\begin{flushleft}
Received: July 17, 2011
\end{flushleft}

\paragraph*{Abstract.} 
This report focuses on the dependencies for the center and radius of the confidence interval that arise when estimating the mass fractal dimensions of anisotropic sampling clusters in the site percolation model.
\vspace{1ex}

\noindent\rule{\textwidth}{.5pt}

\Russian
\begin{flushleft}\bfseries
УДК 519.676
\end{flushleft}
\begin{flushleft}\bfseries\Large
Интервальное оценивание массовой фрактальной размерности для выборки анизотропных перколяционных кластеров
\end{flushleft}
\begin{flushleft}\bfseries
П.В. Москалев (Воронеж, moskalefff@gmail.com), \\
К.В. Гребенников, В.В. Шитов (Воронеж, svw@list.ru)
\end{flushleft}
\begin{flushleft}
Международная конференция 
<<Актуальные проблемы прикладной математики, информатики и механики>>, \\
Воронеж, 19\--21 сентября 2011 г.
\end{flushleft}
\begin{flushleft}
Поступил в Оргкомитет: 17 июля 2011 г.
\end{flushleft}

\paragraph*{Аннотация.}
Настоящий доклад посвящён зависимостям для центра и радиуса доверительного интервала, возникающего при оценивании массовой фрактальной размерности выборки анизотропных кластеров в математической модели перколяции узлов.

\normalsize
\end{titlepage}

В одной из наших работ \cite{moskaleff.2011.03} описан эффект одновременного повышения точности и снижения вычислительной сложности при сокращении размера решётки $x$ для интервального оценивания массовой фрактальной размерности выборки перколяционных кластеров $d_{b1} = d\pm \varepsilon$. Основные идеи методики были изложены в \cite{moskaleff.2011.01}. Речь идёт о модели логарифмически линейной регрессии для векторов относительных суммарных частот $\ln v_i$ узлов кластера, покрываемых элементами текущего размера $\ln r_i$: $\ln v_i = d_{b1} \ln r_i + d_{b0} + e_{bi}$, где сумма квадратов отклонений $\sum e_{bi}^2$ минимизируется по $d_{b0,1}$ методом наименьших квадратов. Кроме того, методика обеспечивает получение репрезентативной оценки $d_{b1}$ во всем диапазоне не только сверх-, но и докритических долей достижимых узлов $p\in (0, 1)$. 

Указанный эффект объясняется тем, что при расположении стартового подмножества узлов кластера в центре квадратной решётки зависимость радиуса интервальной оценки его массовой фрактальной размерности от доли достижимых узлов $\varepsilon(p)$ в окрестности критического значения $p_c = 0,592746\ldots$ имеет локальный минимум $\varepsilon_c$ при $p\to p_c$, окружённый двумя максимумами: докритическим глобальным $\varepsilon_1$ при $p\to p_1 < p_c$ и сверхкритическим локальным $\varepsilon_2$ при $p\to p_2 > p_c$. Соотношение амплитуд максимумов объясняется тем, что малые приращения $p$ в левой окрестности $p_c$ способны вызывать существенно больший прирост выборочной дисперсии $s^2_{d_{b1}}$, чем те же приращения в правой окрестности $p_c$, причём амплитуды обоих максимумов растут с ростом размера решётки $x$. А в непосредственной близости от порога перколяции $p_c$ возникает корреляция размеров реализаций кластеров с размерами области перколяции, которая и минимизирует выборочную дисперсию $s^2_{d_{b1}}$. 

Воспользуемся описанной в \cite{moskaleff.2011.01} методикой для интервального оценивания массовой фрактальной размерности $d_{b1}$ выборки кластеров со стартовым подмножеством узлов, расположенным вдоль нижней границы квадратной перколяционной решётки. Из общих соображений ясно, что формируемые в таких условиях кластеры должны быть анизотропными с фрактальной размерностью: $d_{b1} < 1$ при $p < 1$ и $d_{b1}\to 1$ при $p\to 1$. Но оценка размерности с использованием изотропного (состоящего из квадратов) покрывающего множества приводит к гораздо большим значениям: $1 < d_{b1} < 2$ при $p < 1$ и $d_{b1}\to 2$ при $p\to 1$. Проблема заключается в том, что при линейном стартовом подмножестве рост анизотропного кластера ограничен лишь одним направлением, а изотропное покрывающее множество растёт сразу в двух, что и порождает указанное несоответствие. В таком случае, наиболее логичным решением будет замена базового элемента покрывающего множества с изотропного квадрата, на анизотропный прямоугольник, ориентированный вдоль стартового множества и масштабируемый в направлении роста кластера.

\begin{figure}[hbt]
\centering
\includegraphics[width=.47\linewidth]{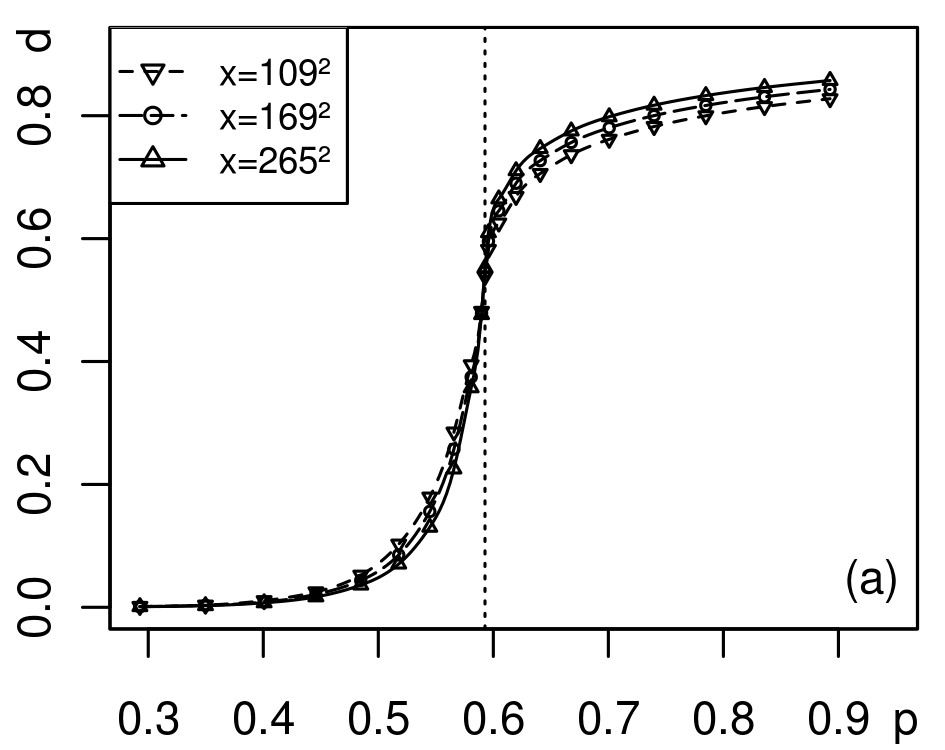}\quad
\includegraphics[width=.47\linewidth]{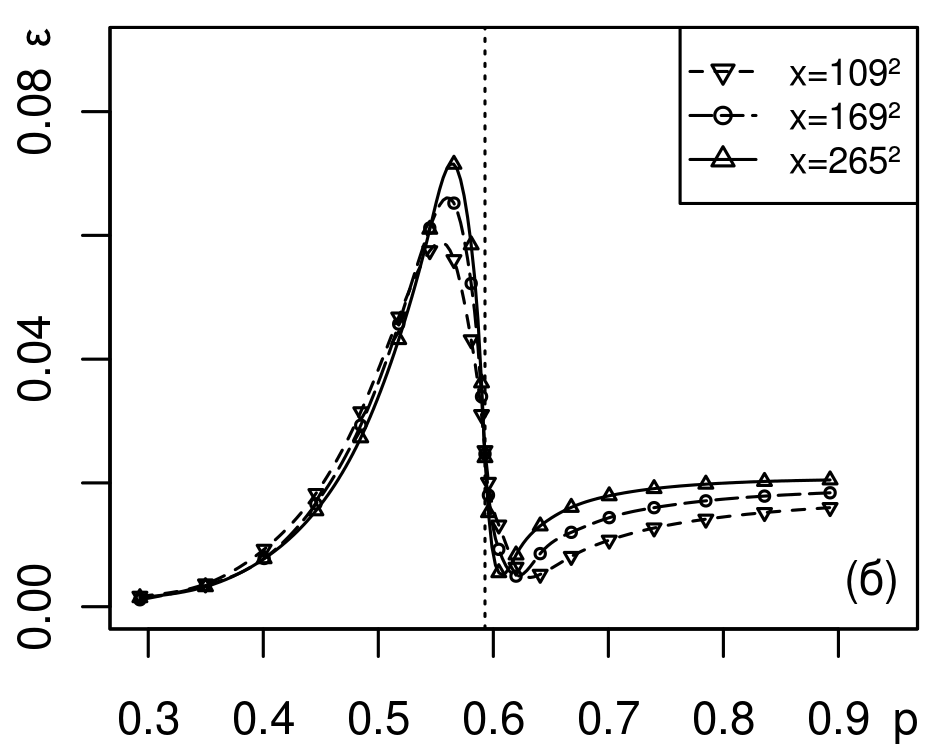}
\caption{Зависимости для центра $d$ и радиуса $\varepsilon$ доверительного интервала массовой фрактальной размерности $\mathbf{I}_{0,95}(d_{b1}) = d \pm \varepsilon$ от доли достижимых узлов $p$ при размерах квадратной решётки $x = 109^2, 169^2, 265^2$ узлов: а)~график $d(p|x)$; б)~график $\varepsilon(p|x)$}
\label{pic:pdex}
\end{figure}

На рис.\,\ref{pic:pdex} показано изменение интервальной оценки массовой фрактальной размерности $\mathbf{I}_{0,95}(d_{b1}) = d\pm \varepsilon$ от доли достижимых узлов $p$ при различных размерах квадратной решётки $x$. Каждая точка на графиках $d(p|x)$ и $\varepsilon(p|x)$ была найдена по выборочной совокупности объёмом 999 реализаций кластеров. Регрессионная модель строилась по 9 точкам, равномерно распределённым в логарифмическом масштабе от характерных размеров покрывающего множества $\ln r_i$. Вертикальная пунктирная линия соответствует критическому значению доли достижимых узлов $p_c$. 

Аналогично изотропному случаю, рассмотренному в \cite{moskaleff.2011.03}, радиус доверительного интервала $\varepsilon(p|x)$ достигает глобального максимума при докритических значениях абсциссы $p_1 < p_c$, причём ордината этого максимума $\varepsilon_1$ также растёт вместе с размером решётки $x$, а отклонение абсциссы максимума от критического значения $(p_c - p_1)$\--- падает. При дальнейшем стремлении $p$ к критическому значению функция $\varepsilon(p|x)$, достигая локального минимума при сверхкритических значениях абсциссы $p_0 > p_c$, а затем снова начинает возрастать, стремясь к правостороннему локальному максимуму при $p_2\to 1$, ордината которого $\varepsilon_2$ также растёт вместе с размером решётки $x$. Примечательно, что с ростом размера решётки $x$ отклонение абсциссы минимума от критического значения $(p_0 - p_c)$ падает, а ордината минимума $\varepsilon_0$\--- судя по графикам не изменяется. Подобно \cite{moskaleff.2011.03} при $p\to 0$ и $p\to p_c$ значимость вариаций от размера решётки $x$ как для центра $d(p|x)$, так и для радиуса доверительного интервала $\varepsilon(p|x)$ массовой фрактальной размерности $d_{b1}$ быстро падает и в пределе при $n\to \infty$ зависимость $d_{b1}(p)$ приобретает кусочно\-/постоянный характер вида: $d_{b1} = 0$ при $p < p_c$, $d_{b1} = 0,456\pm 0,025$ при $p = p_c$ и $d_{b1} = 1$ при $p > p_c$.

Настоящее исследование показывает, что ключевые особенности в поведении интервальной оценки массовой фрактальной размерности $d_{b1}$ сохраняются вне зависимости от изо- или анизотропии стартового и покрывающего множеств. Это в полной мере относится и к выявленному эффекту одновременного повышения точности оценки $d_{b1}$ при снижении вычислительной сложности задачи за счёт сокращения размера решётки $x$ до минимально необходимого уровня, который определяется объёмом выборочной совокупности для модели линейной регрессии в двойных логарифмических координатах $(\ln r_i, \ln v_i)$.

\section*{Список литературы}

\clearpage

\English
\section*{About the authors}

\begin{description}
\item[Moskalev Pavel Valentinovich:] Ph. D., Associate Professor, Department of Mathematics and Theoretical Mechanics, Voronezh State Agricultural University after K.D. Glinki. \\
Tel.: +7 (473) 253-73-71; e-mail: moskalefff@gmail.com

\item[Grebennikov Konstantin Vladimirovich:] Postgraduate Student, Department of Industrial Power Engineering, Voronezh State Technological Academy. \\
Tel.: +7 (473) 255-44-66; e-mail: greb86@mail.ru

\item[Shitov Viktor Vasiljevich:] Doctor of Engineering Science, Full Professor, Head of Department of Industrial Power Engineering, Voronezh State Technological Academy. \\
Tel.: +7 (473) 255-44-66; e-mail: svw@list.ru
\end{description}

\Russian
\section*{Сведения об авторах}

\begin{description}
\item[Москалев Павел Валентинович:] кандидат технических наук, доцент кафедры высшей математики и теоретической механики Воронежского государственного аграрного университета имени К.Д. Глинки. \\
Тел.: +7 (473) 253-73-71; e-mail: moskalefff@gmail.com

\item[Гребенников Константин Владимирович:] аспирант кафедры промышленной энергетики Воронежской государственной технологической академии. \\
Тел.: +7 (473) 255-44-66; e-mail: greb86@mail.ru

\item[Шитов Виктор Васильевич:] доктор технических наук, профессор, заведующий кафедрой промышленной энергетики Воронежской государственной технологической академии. \\
Тел.: +7 (473) 255-44-66; e-mail: svw@list.ru
\end{description}

\end{document}